\documentclass[preprint,preprintnumbers, prd, floatfix, superscriptaddress,nofootinbib] {revtex4-1}
\usepackage{epsfig}
\usepackage{subfigure}
\usepackage{dcolumn}
\usepackage{bm}
\usepackage[usenames ,dvipsnames]{xcolor}
\usepackage{slashed}
\usepackage{graphicx,color}

\begin{document}
\title{Semileptonic $B^{-}\to\pi^{+}\pi^{-}\ell^{-}\bar\nu_\ell$ decay\\
with $\pi\pi$ invariant mass spectrum}
\author{Shang-Yuu Tsai}
\email{shangyuu@gmail.com}
\affiliation{School of Physics and Information Engineering, Shanxi Normal University, Linfen 041004, China}

\author{Yu-Kuo Hsiao}
\email{yukuohsiao@gmail.com}
\affiliation{School of Physics and Information Engineering, Shanxi Normal University, Linfen 041004, China}

\date{\today}

\begin{abstract}
BELLE has recently reported the measurement of the branching fraction of
the semileptonic $B^{-}\to\pi^{+}\pi^{-}\ell^{-}\bar\nu_\ell$ decay,
where $\ell$ represents an electron or a muon.
With the new information on the $\pi\pi$ invariant mass spectrum, 
we extract $|V_{ub}|=(3.31\pm 0.61)\times 10^{-3}$
in agreement with those from the other exclusive $B$ decays.
In particular,
we determine the non-resonant $B\to\pi\pi$ transition form factors, 
and predict the non-resonant branching fraction
${\cal B}(B^{-}\to\pi^{+}\pi^{-}\ell^{-}\bar\nu_\ell)=(3.5\pm 1.4^{+4.3}_{-2.4})\times 10^{-5}$,
which is accessible to the BELLEII and LHCb experiments.
\end{abstract}

\maketitle

\section{Introduction}
For the Cabibbo-Kobayashi-Maskawa (CKM) matrix element $|V_{ub}|$, 
there have been the long-standing inconsistent determinations 
from the inclusive and exclusive $b$-hadron decays~\cite{review,pdg},
which might indicate the existence of new physics~\cite{Kang:2013jaa,
Feldmann:2015xsa,Crivellin:2009sd,Buras:2010pz,Crivellin:2014zpa,
Hsiao:2017umx,Hsiao:2015mca,Hsiao:2018zqd,Kim:2017dfr}.
For a careful examination, 
the exclusive $B^{-}\to\pi^{+}\pi^{-}\ell^{-}\bar\nu_\ell$ decay
can provide another path to determining $|V_{ub}|$, where
$\ell$ represents an electron or a muon. 
Nonetheless, although $B^{-}\to\pi^{+}\pi^{-}\ell^{-}\bar\nu_\ell$ 
has been observed many times~\cite{Behrens:1999vv,
Hokuue:2006nr,Sibidanov:2013rkk,delAmoSanchez:2010af},
it is essentially $B^{-}\to\rho^0 \ell^{-}\bar\nu_\ell$ along with $\rho^0\to\pi^+\pi^-$,
instead of a genuine four-body decay.

Recently, BELLE has newly reported the measurement of the branching fractions of
$B^{-}\to\pi^{+}\pi^{-}\ell^{-}\bar\nu_\ell$
with the full $\pi\pi$ invariant mass ($M_{\pi\pi}$) spectrum~\cite{Beleno:2020gzt}.
In addition to the resonant processes of $B^{-}\to R\ell^{-}\bar\nu_\ell,R\to\pi^{+}\pi^{-}$
with $R=\rho^0$ and $f_2\equiv f_2(1270)$, the non-resonant contribution is also found. 
Explicitly, we present the branching fractions as~\cite{Beleno:2020gzt,pdg,thesis}
\begin{eqnarray}\label{data1}
&&{\cal B}_{\text{T}}(B^{-}\to\pi^{+}\pi^{-}\ell^{-}\bar\nu_\ell)
=(22.7^{+1.9}_{-1.6}\pm 3.4)\times 10^{-5}
\,,\nonumber\\
&&{\cal B}_{\rho}(B^{-}\to\rho^0\ell^{-}\bar\nu_\ell,\rho^0\to\pi^{+}\pi^{-})
=(15.8\pm 1.1)\times 10^{-5}
\,,\nonumber\\
&&{\cal B}_{f_2}(B^{-}\to f_2\ell^{-}\bar\nu_\ell,f_2\to\pi^{+}\pi^{-})
=(1.8\pm 0.9^{+0.2}_{-0.1})\times 10^{-5}
\,,\nonumber\\
&&{\cal B}_{\text{N}}(B^{-}\to\pi^{+}\pi^{-}\ell^{-}\bar\nu_\ell)
=(5.1\pm 4.3)\times 10^{-5}\,,
\end{eqnarray}
where ${\cal B}_{\text{T,N}}$
denote the total and non-resonant branching fractions, respectively,
while ${\cal B}_{\rho}\simeq{\cal B}(B^{-}\to\rho^0\ell^{-}\bar\nu_\ell)\times{\cal B}(\rho^0\to\pi^+\pi^-)$
is from PDG~\cite{pdg}.
By excluding ${\cal B}_{\rho,f_2}$ from ${\cal B}_{\text{T}}$,
we estimate ${\cal B}_{\text{N}}$ in Eq.~(\ref{data1}).

\begin{figure}[t]
\centering
\includegraphics[width=3.2in]{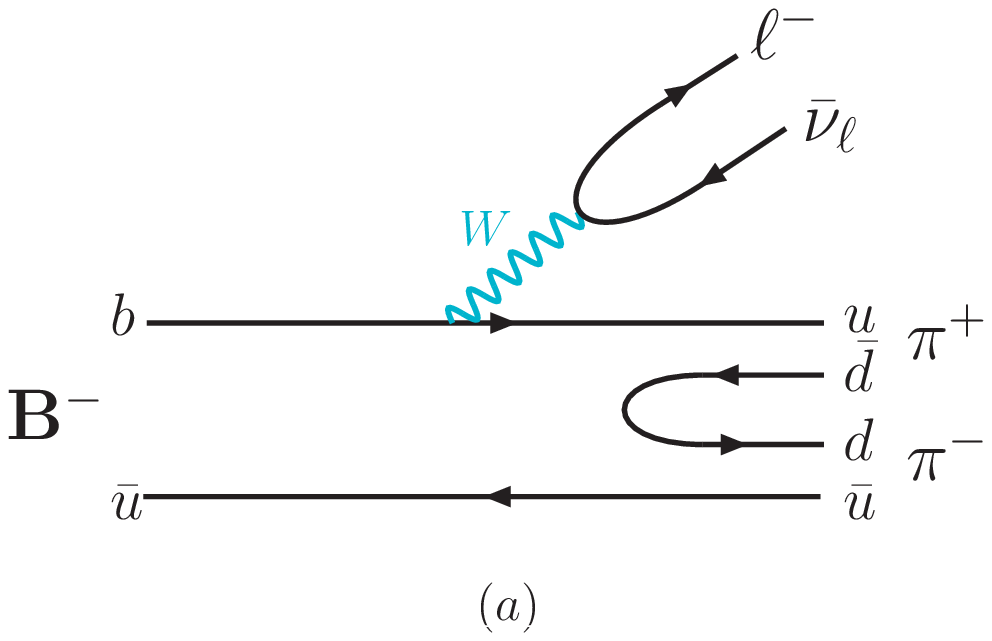}
\includegraphics[width=3.2in]{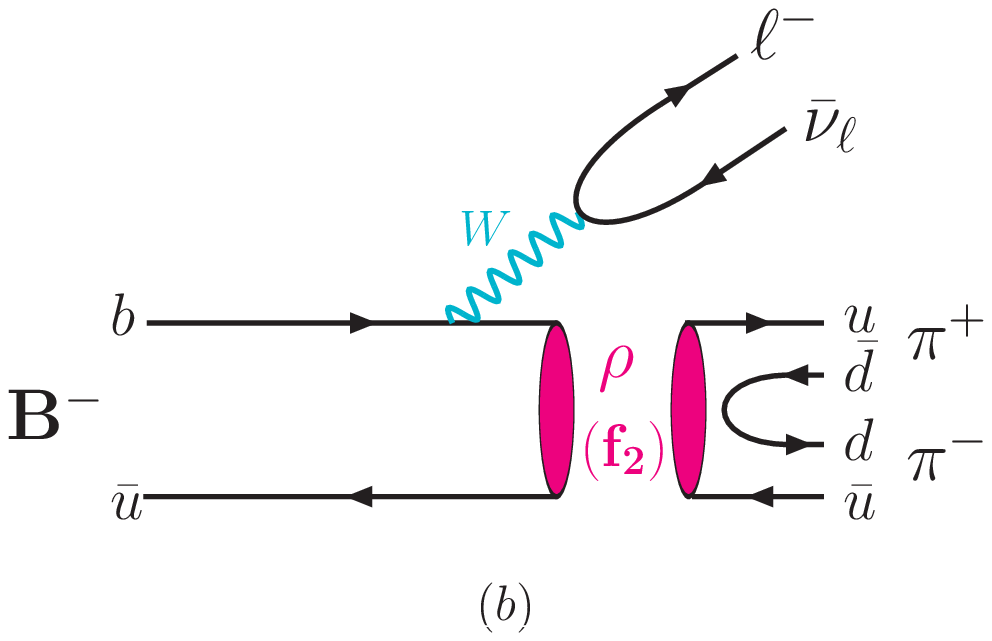}
\caption{$B^{-}\to\pi^{+}\pi^{-}\ell^{-}\bar\nu_\ell$ with
(a) non-resonant and (b) resonant contributions.}\label{semi}
\end{figure}

As depicted in Fig.~\ref{semi}, $B^{-}\to\pi^{+}\pi^{-}\ell^{-}\bar\nu_\ell$ 
proceeds through the resonant and non-resonant $B\to \pi\pi$ transitions, respectively,
with the lepton-pair produced from the emitted $W$-boson.
One has been enabled to parameterize
the resonant $B\to\rho(f_2),\rho(f_2)\to\pi\pi$ transition~\cite{Cheng:2020ipp}.
Despite the theoretical attempts~\cite{Kang:2013jaa,Lee:1992ih,
Fajfer:1998yc,Hsiao:2017nga,Chua:2002pi,Chua:2004mi,
Boer:2016iez,Cheng:2019hpq,Feldmann:2018kqr,
Cheng:2017smj,Hambrock:2015aor,Hsiao:2019ann},
the non-resonant $B\to\pi\pi$ transition is still poorly understood.
With the full $\pi\pi$ invariant mass spectrum provided
for the first time, the information on the non-resonant $B\to\pi\pi$ transition form factors
($F_{\pi\pi}$) becomes available. Hence, we propose to
newly extract $|V_{ub}|$ and $F_{\pi\pi}$, by which we will be able to study ${\cal B}_{\text{N}}$. 
We will also study the angular distribution and its asymmetry
to be compared to the future measurements.
%
%
\section{Theoretical Framework}
The semileptonic $B^-\to\pi^+\pi^-\ell^-\bar \nu_\ell$ decay
is observed with the full $M_{\pi\pi}$ spectrum,
which indicates the existence of the non-resonant contribution~\cite{Beleno:2020gzt}.
Moreover,
the simulation is performed to seek the resonances
that contribute to $B^-\to\pi^+\pi^-\ell^-\bar \nu_\ell$.
It turns out that only a dominant peak and a small bump
are observed, which correspond to $B^-\to \rho^0\ell\bar \nu,f_2\ell\bar \nu$,
respectively, with $\rho^0,f_2\to \pi^+\pi^-$.
Therefore, the total amplitude of $B^-\to \pi^+\pi^-\ell^-\bar \nu_\ell$
can be written as
\begin{eqnarray}\label{amp1}
{\cal M}_{\text T}
&=&{\cal M}_{\text N}(B^-\to \pi^+\pi^-\ell\bar \nu_\ell)
+{\cal M}_{\rho}(B^-\to \rho^0\ell^-\bar \nu_\ell,\rho^0\to \pi^+\pi^-)\nonumber\\
&+&{\cal M}_{f_2}(B^-\to f_2\ell^-\bar \nu_\ell,f_2\to \pi^+\pi^-)\,,\nonumber\\
{\cal M}_{\text {N(R)}}
&=&\frac{G_F V_{ub}}{\sqrt 2}\langle \pi^+\pi^-|\bar u\gamma_\mu(1-\gamma_5)b|B^-\rangle_{\text{N(R)}}\,
\bar u_\ell\gamma^\mu(1-\gamma_5)v_\nu\,,
\end{eqnarray}
with $R=(\rho,f_2)$. The matrix elements of the (non-)resonant $B$ meson to $\pi\pi$ transitions
can be parameterized as~\cite{Pais:1968zza,Lee:1992ih}
\begin{eqnarray}\label{ff1}
&&\langle \pi^+(p_a)\pi^-(p_b)|\bar u\gamma_\mu(1-\gamma_5)b|B^-\rangle_N\nonumber\\
&=&
h\epsilon_{\mu\nu\alpha\beta}p_B^\nu p^\alpha (p_b-p_a)^\beta
+irq_\mu+iw_+ p_\mu+iw_-(p_b-p_a)\,,\nonumber\\
&&\langle \pi^+(p_a)\pi^-(p_b)|\bar u\gamma_\mu(1-\gamma_5)b|B^-\rangle_{\rho(f_2)}\nonumber\\
&=&
\langle \pi^+\pi^-|\rho(f_2)\rangle \frac{i}{(t-m_{\rho(f_2)}^2)+im_{\rho(f_2)}\Gamma_{\rho(f_2)}}
\langle \rho(f_2)|\bar u\gamma_\mu(1-\gamma_5)b|B^-\rangle\,,
\end{eqnarray}
with $p=p_b+p_a$, $q=p_B-p=p_\ell+p_\nu$, $(s,t)\equiv (q^2,p^2)$, and
the form factors $F_{\pi\pi}=(h, r, w_{\pm})$.
The matrix elements of $B\to \rho(f_2)$ transition
are written as~\cite{Wang:2010ni,Cheng:2010yd,Zuo:2021kui}\\
\begin{eqnarray}\label{ff2}
\langle \rho(f_2)|\bar u\gamma_\mu b|B\rangle&=&\epsilon_{\mu\nu\alpha\beta}
\epsilon^{(\prime)\nu}p_B^{\alpha}p_{\rho(f_2)}^{\beta}
\frac{2V_1^{(\prime)}}{m_{B}+m_{\rho(f_2)}}\;,\nonumber\\
\langle \rho(f_2)|\bar u\gamma_\mu \gamma_5 b|B\rangle
&=&i\bigg[\epsilon^{(\prime)}_\mu
-\frac{\epsilon^{(\prime)}\cdot p_B}{s}q_\mu\bigg](m_B+m_{\rho(f_2)})A_1^{(\prime)}
+ i\frac{\epsilon^{(\prime)}\cdot p_B}{s}q_\mu(2m_{\rho(f_2)})A_0^{(\prime)}\nonumber\\
&-&i\bigg[(p_B+p_{\rho(f_2)})_\mu-
\frac{m^2_B-m^2_{\rho(f_2)}}{s}q_\mu \bigg](\epsilon^{(\prime)}\cdot p_B)
\frac{A_2^{(\prime)}}{m_B+m_{\rho(f_2)}}\;,
\end{eqnarray}
with $\epsilon^{\prime\mu}\equiv \epsilon^{\mu\nu}p_{B\nu}/m_B$ and
the form factors $F_{\rho(f_2)}=(V_1^{(\prime)}, A_{0,1,2}^{(\prime)})$, where
$\epsilon^\nu$ and $\epsilon^{\mu\nu}$ are the polarization vector and tensor, respectively. 
To describe the $\rho^0,f_2\to \pi^+\pi^-$ decays, 
$\langle \pi\pi|\rho,f_2\rangle$ in Eq.~(\ref{amp1}) 
are given by~\cite{Cheng:2020ipp,Kim:2017dfr,Suzuki:1993zs}
\begin{eqnarray}
\langle \pi\pi|\rho\rangle&=&g_1 \epsilon\cdot (p_b-p_a)\,,\;\nonumber\\
\langle \pi\pi|f_2\rangle&=&g_2\epsilon^{\mu\nu}p_{a\mu}p_{b\nu}\,,
\end{eqnarray}
where $g_{1,2}$ are strong coupling constants. To sum over 
the vector and tensor spins for $\rho$ and $f_2$, respectively,
as the intermediate states in the resonant $B\to\pi\pi$ transitions,
we use the following identities~\cite{Wang:2010ni,Cheng:2010yd,Zuo:2021kui},
\begin{eqnarray}
\Sigma\epsilon_\mu \epsilon^*_{\mu'}
&=&M_{\mu\mu'}\,,\nonumber\\ 
\Sigma\epsilon_{\mu\nu} \epsilon^*_{\mu'\nu'}
&=&{1\over 2}M_{\mu\mu'}M_{\nu\nu'}+{1\over 2}M_{\mu\nu'}M_{\nu\mu'}
-{1\over 3}M_{\mu\nu}M_{\mu'\nu'}\,,
\end{eqnarray}
with $M_{\mu\mu'}=-g_{\mu\mu'}+p_\mu p_{\mu'}/p^2$.
The form factors in Eqs.~(\ref{ff1},\,\ref{ff2}) are momentum-dependent, modelled 
in the single-pole or double-pole forms~\cite{Wang:2010ni,Cheng:2010yd,Zuo:2021kui}:
\begin{eqnarray}\label{p_depent}
F_\rho(s)&=&\frac{F_\rho(0)}{1-s/m_V^2}\,,\nonumber\\
F_{f_2}(s)&=&\frac{F_{f_2}(0)}{(1-s/m_B^2)^2}\,,\nonumber\\
F_{\pi\pi}(t)&=&\frac{F_{\pi\pi}(0)}{1-a\,(t/m_B^2)+b\,(t/m_B^2)^2}\,,
\end{eqnarray}
where $F_{\rho,f_2}(s)$ have been studied in QCD models, whereas
$(a,b,F_{\pi\pi}(0))$ need to be extracted in the global fit.

\begin{figure}[t]
\centering
\includegraphics[width=3.0 in]{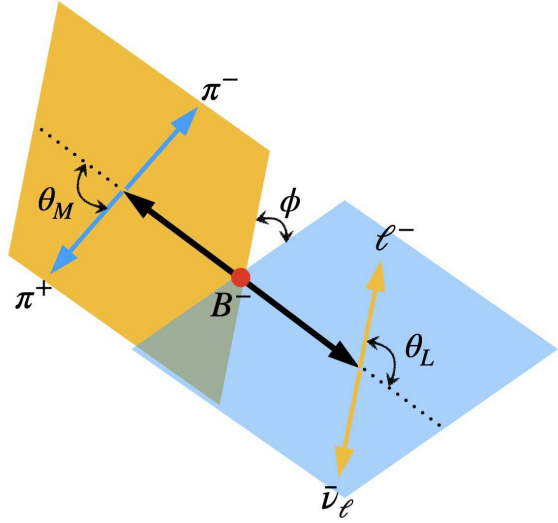}
\caption{The angular variables $(\theta_{M}, \theta_{L},\phi)$
in the four-body $B^-\to\pi^+\pi^-\ell^-\bar \nu_\ell$ decay.}\label{4body}
\end{figure}

For the four-body decay channel $B^-(p_B)\to \pi^+(p_a)\pi^-(p_b)\ell^-(p_\ell)\bar \nu_\ell(p_\nu)$,
one has to integrate over
the kinematic variables $(s,t,\theta_M,\theta_L,\phi)$ in the phase space.
See Fig.~\ref{4body},
$\theta_{M(L)}$ is the angle between $\pi^+$ and $\pi^-$ ($\ell^-$ and $\bar \nu_\ell$)
moving directions in the $\pi^+\pi^-$ ($\ell^-\bar \nu_\ell$) rest frame.
In addition,
the angle $\phi$ is between the $\pi^+\pi^-$ and $\ell^-\bar \nu_\ell$ planes,
defined by $\vec{p}_{a,b}$ and $\vec{p}_{\ell,\bar \nu_\ell}$, respectively,
in the $B$-meson rest frame.
Then, the partial decay width reads~\cite{Geng:2011tr,Geng:2012qn}
\begin{eqnarray}\label{dGamma}
d\Gamma=\frac{|{\cal M}|^2}{4(4\pi)^6 m_B^3}X
\alpha_M\alpha_L\, ds\, dt\, d\text{cos}\,\theta_M\, d\text{cos}\,\theta_L\, d\phi\,,
\end{eqnarray}
where $X$, $\alpha_{M}$ and $\alpha_{L}$ are defined by
\begin{eqnarray}
X&=&\bigg[\frac{1}{4}(m_B^2-s-t)^2-st\bigg]^{1/2}\,,\nonumber\\
\alpha_{M}&=&\frac{1}{t}\lambda^{1/2}(t,m_{\pi}^2,m_{\pi}^2)\,,\nonumber\\
\alpha_{L}&=&\frac{1}{s}\lambda^{1/2}(s,m_{\ell}^2,m_{\bar \nu}^2)\,,
\end{eqnarray}
with $\lambda(a,b,c)=a^2+b^2+c^2-2ab-2bc-2ca$.
The allowed ranges for $(s,t)$ and the angular variables
($\theta_M,\theta_L,\phi$) are given by
\begin{eqnarray}
(m_{\ell}+m_{\bar \nu_\ell})^2&\leq& s\leq (m_{B}-\sqrt{t})^2\,,\;\;\nonumber\\
4m_\pi^2&\leq& t\leq (m_B-m_{\ell}-m_{\bar \nu_{\ell}})^2\,,\nonumber\\
0&\leq& \theta_{M,L}\leq \pi\,,\;\;\nonumber\\
0&\leq& \phi\leq 2\pi\,,
\end{eqnarray}
with $m_{\ell}+m_{\bar \nu_\ell}\simeq 0$.
From Eq. (\ref{dGamma}), we define the angular distribution asymmetry as
\begin{eqnarray}\label{AFB}
{A}_{\theta_M}\equiv\frac{\int^{+1}_0\frac{d\Gamma}{d\cos\theta_{M}}d\cos\theta_M
-\int^0_{-1}\frac{d\Gamma}{d\cos\theta_M}d\cos\theta_M}{\int^{+1}_0\frac{d\Gamma}{d\cos\theta_M}
d\cos\theta_M+\int^0_{-1}\frac{d\Gamma}{d\cos\theta_M}d\cos\theta_M}\;,
\end{eqnarray}
where $d\Gamma/d\cos\theta_M$ is the angular distribution.
%
\begin{table}[t!]
\caption{The $B$ to $(\rho, f_2)$ transition form factors
with $M_V=7.0$~GeV in Eq.~(\ref{p_depent})~\cite{DelDebbio:1997ite,Cheng:2010yd}. 
Here, we present $\sqrt 2 F_{\rho^0}=F_{\rho}$ for the $B$ to $\rho^0$ transition.}\label{tab1}
\begin{tabular}{|c|ccc|}
\hline
&$V_1^{(\prime)}$& $A_1^{(\prime)}$ & $A_2^{(\prime)}$ \\
\hline
$\sqrt 2 F_{\rho^0}(0)$
&$0.35^{+0.06}_{-0.05}$ & $0.27^{+0.05}_{-0.04}$ &  $0.26^{+0.05}_{-0.03}$ \\
$F_{f_2}(0)$
&$(0.18\pm 0.02)$ & $(0.13\pm 0.02)$ &  $(0.12\pm 0.02)$\\
\hline
\end{tabular}
\end{table}
%
\newpage
\section{Numerical analysis}
\begin{figure}[t!]
\centering
\includegraphics[width=3.8in]{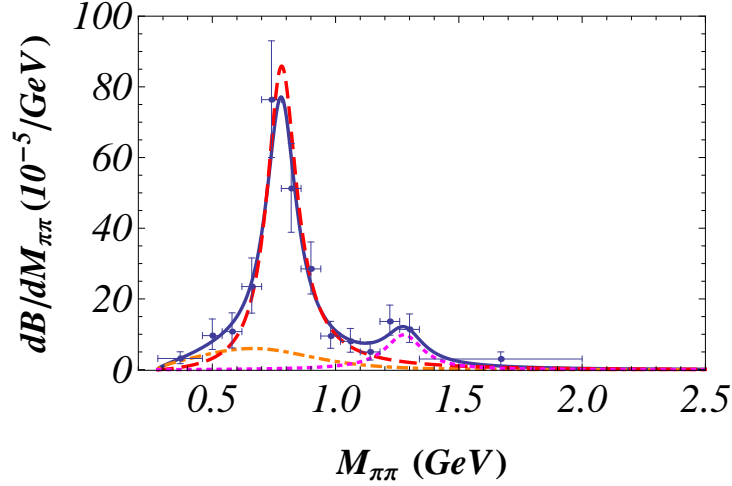}
\caption{The $\pi\pi$ invariant mass spectrum, where
the solid curve that takes into account the all contributions
explains the data points from BELLE~\cite{Beleno:2020gzt}.
On the other hand, the dashed (dotted) and dot-dashed curves depict
the contributions from $B^-\to\rho(f_2)\ell\bar \nu,\rho(f_2)\to \pi^+\pi^-$, and 
non-resonant $B^{-}\to\pi^{+}\pi^{-}\ell^{-}\bar\nu_\ell$, respectively.}\label{fig:spectrum}
\end{figure}

In the numerical analysis, 
we perform the minimum $\chi^2$-fit, in order to extract 
$|V_{ub}|$, $F_{\pi\pi}$ and $\delta_{1,2}$ as the free parameters,
where $\delta_{1(2)}$ is the relative phase for ${\cal A}_{\rho(f_2)}$.
The equation of the $\chi^2$-fit is given by
\begin{eqnarray}\label{fitEQ}
\chi^2&=&
\bigg(\frac{{\cal B}_{\rho\,th}-{\cal B}_{\rho\,ex}}{\sigma_{\rho\,ex}}\bigg)^2+
\bigg(\frac{{\cal B}_{f_2\,th}-{\cal B}_{f_2\,ex}}{\sigma_{f_2\,ex}}\bigg)^2
\,\nonumber\\
&&+
\sum_{i} \bigg(\frac{\frac{d{\cal B}^i_{th}}{dM_{\pi\pi}}-\frac{d{\cal B}^i_{ex}}{dM_{\pi\pi}}}{\sigma_{ex}^i}\bigg)^2
+\sum_{j} \bigg(\frac{F_{\rho(f_2)}^j-F_{th\,\rho(f_2)}^j}{\delta F_{th\,\rho(f_2)}^j}\bigg)^2\,,
\end{eqnarray}
where $d{\cal B}/dM_{\pi\pi}$ denotes the partial branching ratio, and
$\sigma_{ex}$ ($\delta F_{th}$) the uncertainty from the observation (form factor).
${\cal B}_{\rho(f_2)\,th}$ and $d{\cal B}_{th}/dM_{\pi\pi}$ 
are the theoretical inputs from the amplitudes in Eq.~(\ref{amp1}), and
the experimental inputs are given in Eq.~(\ref{data1}) and Fig.~\ref{fig:spectrum}.
We take $F_\rho$ and $F_{f_2}$ in Table~\ref{tab1} as the initial values in Eq.~(\ref{fitEQ}), 
together with $|g_1|=5.98$ and $|g_2|=18.56$~GeV$^{-1}$~\cite{Hsiao:2019ait,Suzuki:1993zs}.

Subsequently, we extract that
\begin{eqnarray}\label{fit1}
&&
|V_{ub}|=(3.31\pm 0.61)\times 10^{-3}\,,\nonumber\\
&&
a=(0.96\pm 0.93)\times m_B^2,\;
b=(1.84\pm 0.87)\times m_B^4,\;\nonumber\\
&&
h(0)=1.90\pm 0.43\,,\,
w_+(0)=6.16\pm 3.41\,,\,
w_-(0)= 3.67\pm 1.79\,,\nonumber\\
&&
(\delta_1,\delta_2)=(-111.6\pm 29.3,0.0\pm 1.4)^\circ\,\nonumber\\
&&
\chi^2/n.d.f=1.1\,,
\end{eqnarray}
with $n.d.f=7$ the number of degrees of freedom. 
The form factors $V_1^{(\prime)}$ and $A_{1,2}^{(\prime)}$ 
are fitted to slightly deviate from their initial inputs in Table~\ref{tab1},
given by
\begin{eqnarray}\label{fit2}
&&
(V_1(0),A_1(0),A_2(0))=(0.35\pm 0.06,0.29\pm 0.04,0.28\pm 0.04)\,,\nonumber\\
&&
(V'_1(0),A'_1(0),A'_2(0))=(0.18\pm 0.02,0.11\pm 0.02,0.14\pm 0.02)\,.
\end{eqnarray}
Nonetheless, $r$ and $A_0^{(\prime)}$ in Eqs.~(\ref{ff1}, \ref{ff2}) 
are not involved in the global fit, since they have been vanishing 
with $q_\mu\bar u_\ell\gamma^\mu(1-\gamma_5)v_\nu=0$
in the amplitudes, where the lepton pair is nearly massless. 

Using the fit results in Eqs.~(\ref{fit1},\ref{fit2}), we obtain
\begin{eqnarray}\label{result1}
&&{\cal B}_{\text{T}}(B^{-}\to\pi^{+}\pi^{-}\ell^{-}\bar\nu_\ell)
=(19.6\pm 7.9^{+7.5+0.7}_{-5.4-0.1})\times 10^{-5}\,,\nonumber\\
&&{\cal B}_{\rho}(B^{-}\to\rho^0\ell^{-}\bar\nu_\ell,\rho^0\to\pi^{+}\pi^{-})
=(15.8\pm 6.4^{+7.1}_{-5.7})\times 10^{-5}\,,\nonumber\\
&&{\cal B}_{f_2}(B^{-}\to f_2\ell^{-}\bar\nu_\ell,f_2\to\pi^{+}\pi^{-})
=(2.6\pm 1.1^{+1.2}_{-0.9})\times 10^{-5}\,,\nonumber\\
&&{\cal B}_{\text{N}}(B^{-}\to\pi^{+}\pi^{-}\ell^{-}\bar\nu_\ell)
=(3.5\pm 1.4^{+4.3}_{-2.4})\times 10^{-5}\,,
\end{eqnarray}
where the first errors are from $|V_{ub}|$,
the second ones from the form factors, and
the third error for ${\cal B}_T$ from the relative phase $\delta_1$.
Moreover, we draw the partial branching fractions as the functions of
$M_{\pi\pi}$ and $\cos\theta_M$ in Fig.~\ref{fig:spectrum} and Fig.~\ref{AD}, respectively.
We also calculate the angular distribution asymmetries,
given by
\begin{eqnarray}\label{result2}
&&
{A}_{\theta_M,\text{T}}(B^{-}\to\pi^{+}\pi^{-}\ell^{-}\bar\nu_\ell)
=(1.3\pm 8.9^{+0.8}_{-2.5})\%\,,\nonumber\\
&&
{A}_{\theta_M,\rho}(B^{-}\to\rho^0\ell^{-}\bar\nu_\ell,\rho^0\to\pi^{+}\pi^{-})
=(0.20\pm 0.04)\%\,,\nonumber\\
&&
{A}_{\theta_M,f_2}(B^{-}\to f_2\ell^{-}\bar\nu_\ell,f_2\to\pi^{+}\pi^{-})
=(0.31\pm 0.08)\%\,,\nonumber\\
&&
{A}_{\theta_M,\text{N}}(B^{-}\to\pi^{+}\pi^{-}\ell^{-}\bar\nu_\ell)
=(-43.0\pm 22.3)\%\,,
\end{eqnarray}
where the first errors come from the uncertainties of the form factors, and
the second error for ${A}_{\theta_M,\text{T}}$ is from the relative phase $\delta_1$.
%

\section{Discussions and Conclusions}
%
\begin{figure}[t!]
\centering
\includegraphics[width=3.8in]{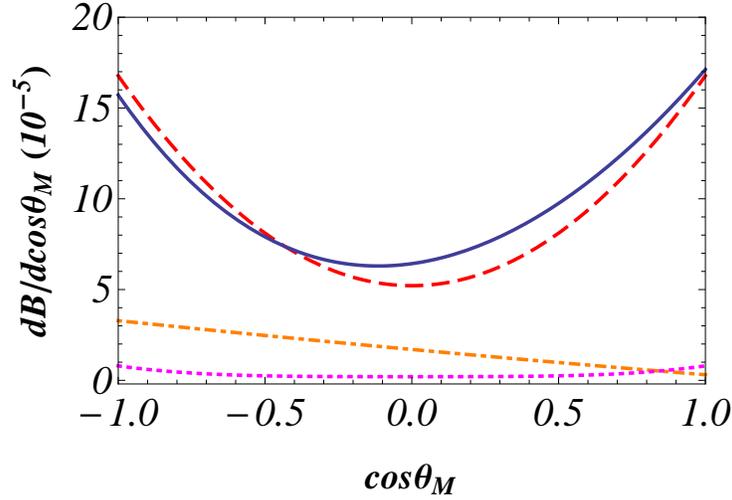}
\caption{Angular distributions of $B^{-}\to\pi^{+}\pi^{-}\ell^{-}\bar\nu_\ell$,
where the solid, dashed, dotted and dot-dashed curves 
represent the same contributions as those in Fig.~\ref{fig:spectrum}.}\label{AD}
\end{figure}
%
We study $B^-\to\pi^+\pi^-\ell\bar \nu$, in order to explain
the $\pi\pi$ invariant mass spectrum observed by BELLE~\cite{Beleno:2020gzt}.
In Fig.~\ref{fig:spectrum}, 
the curves for $B^-\to(\rho^0,f_2)\ell\bar \nu,(\rho^0,f_2)\to \pi^+\pi^-$
are shown to barely fit the first three data points in the spectrum. 
Nonetheless, the non-resonant $B^-\to\pi^+\pi^-\ell\bar \nu$
raises the contribution as the dot-dashed curve describes. As a result, 
the solid curve that takes into account the resonant and non-resonant contributions
is able to explain the data, with $\chi^2/d.o.f=1.1$ that presents a reasonable fit.
The relative phase
$\delta_1=-111.6^\circ$ causes a destructive interference
between the non-resonant $B^-\to \pi^+\pi^-\ell\bar \nu$ and
$B^-\to\rho^0\ell\bar \nu,\rho^0\to\pi^+\pi^-$.
As a demonstration, we turn off $\delta_1$ and obtain ${\cal B}_T=22.2\times 10^{-5}$.
By contrast, $\delta_2$ is fitted to be zero, in accordance with
the fact that the non-resonant contribution is tiny in the range of $M_{\pi\pi}>1$~GeV,
barely having the interference with $B^-\to f_2\ell\bar \nu,f_2\to\pi^+\pi^-$.

It turns out that ${\cal B}_{\text{N}}=(3.5\pm 1.4^{+4.3}_{-2.4})\times 10^{-5}$
is given for the first time.
Also importantly, we determine $|V_{ub}|=(3.31\pm 0.61)\times 10^{-3}$
from the first genuine four-body semileptonic $B\to\pi\pi\ell\bar \nu$ decay,
instead of $B^-\to \rho^0\ell\bar \nu,\rho^0\to\pi^+\pi^-$.
For the angular distribution asymmetries, we obtain
${A}_{\theta_M,\rho(f_2)}=0$, showing the symmetric distributions as 
the curves in Fig.~\ref{AD}.
By contrast, $|{A}_{\theta_M,\text{N}}|$ is as large as $40\%$.
This is due to the main contributions from the form factors
$w_+(p_b+p_a)_\mu$ and $w_-(p_b-p_a)$. 
With $p_b+p_a=(2E_b,\vec{0})$ and $p_b-p_a=(0,2\vec{p}_b)$
in the $\pi^+(p_a)\pi^-(p_b)$ rest frame (see Fig.~\ref{4body}), 
the projection of $w_\mp(p_b\mp p_a)$ onto the four-momentum of the lepton pair system
causes a $\cos\theta_M$-(in)dependent term,
such that their interference leads to the large angular distribution asymmetry.

In summary, we have studied the semileptonic $B^-\to\pi^+\pi^-\ell\bar \nu$ decay.
With the full $\pi\pi$ invariant mass spectrum observed by BELLE,
we have determined $|V_{ub}|=(3.31\pm 0.61)\times 10^{-3}$ agreeing with
the other exclusive determinations. Besides, we have extracted 
the non-resonant $B\to\pi\pi$ transition form factors, by which
we have predicted the non-resonant branching fraction
${\cal B}_{\text{N}}(B^{-}\to\pi^{+}\pi^{-}\ell^{-}\bar\nu_\ell)
=(3.5\pm 1.4^{+4.3}_{-2.4})\times 10^{-5}$. We have also predicted
the non-resonant angular distribution asymmetry
${A}_{\theta_M,\text{N}}(B^{-}\to\pi^{+}\pi^{-}\ell^{-}\bar\nu_\ell)
=(-43.0\pm 22.3)\%$ to be checked by the future measurements.

\section*{ACKNOWLEDGMENTS}
This work was supported by National Science Foundation of China (No. 11675030).

\end{document}